\documentclass[iop,apj]{emulateapj}

\newcommand \cmg                {\,cm$^2$ g$^{-1}$}
\newcommand \gcm                {\,g cm$^{-2}$}

\begin{document}
   \title{Evidence of fast pebble growth near condensation fronts \\
   in the HL Tau protoplanetary disk} 

\author{Ke Zhang\altaffilmark{1},  Geoffrey A. Blake\altaffilmark{2}, Edwin A. Bergin\altaffilmark{3}}

\altaffiltext{1}{Division of Physics, Mathematics \& Astronomy, MC 249-17, California Institute of Technology, Pasadena, CA 91125, USA; kzhang@astro.caltech.edu}
\altaffiltext{2}{Division of Geological \& Planetary Sciences, MC 150-21, California Institute of Technology, Pasadena, CA 91125, USA}
\altaffiltext{3}{Department of Astronomy, University of Michigan, 500 Church Street, Ann Arbor, Michigan 48109, USA}

  \begin{abstract}
Water and simple organic molecular ices dominate the mass of solid materials available for planetesimal and planet formation beyond the water snow line. Here we analyze ALMA long baseline 2.9, 1.3 and 0.87\,mm continuum images of the young star HL Tau, and suggest that the emission dips observed are due to rapid pebble growth around the condensation fronts of abundant volatile species. Specifically, we show that the prominent innermost dip at 13\,AU is spatially resolved in the 0.87\,mm image, and its center radius is coincident with the expected mid-plane condensation front of water ice. In addition, two other prominent dips, at distances of 32 and 63 AU, cover the mid-plane condensation fronts of  pure ammonia or ammonia hydrates and clathrate hydrates (especially with 
CO and N$_2$) formed from amorphous water ice.  The spectral index map of HL Tau between 1.3 and 0.87\,mm shows that the flux ratios inside the dips are statistically larger than those of nearby regions in the disk.  This variation can be explained by a model with two dust populations, where most of solid mass resides in a component that has grown into decimeter size scales inside the dips. Such growth is in accord with recent numerical simulations of volatile condensation, dust coagulation and settling.
 \end{abstract}


   \maketitle


\section{Introduction}
\label{sec:intro}

The frost lines of abundant ices have long been thought to be important for planet formation, in part because the extra material provided by condensation can significantly enhance the local mass surface density of solids in the disk \citep{Hayashi81, Stevenson88, Desch07}. This in turn significantly shortens the time scale needed to form protoplanets in core-accretion models \citep{Pollack96}.  Furthermore, the bulk composition of terrestrial planets and the cores of gas or ice giants is determined by that of the planetesimals from which they are built. Because volatility is highly species specific, 
it is the condensation fronts of the principal solid-vapor reservoirs in the mid-plane of protoplanetary disks that set the rough boundaries of planetesimals with different compositions \citep{Oberg11}.

Recent experiments and numerical simulations of dust coagulation and settling suggest that condensation fronts may play a crucial role in planetesimal formation via two aspects: First, the growth efficiency of dust grains is composition dependent. Lab experiments show that icy aggregates are significantly more `sticky', and resistant to compaction than silicate aggregates \citep{Guttler10, Seizinger13, Kelling14}.
The second, perhaps more important, aspect is that the enhanced local surface mass density near snow lines may produce pressure bumps, high dust-to-gas mass ratios, or viscosity gradients, thus triggering instabilities that form planetesimals \citep{Kretke07, Brauer08, Ros13, Bitsch14,Drazkowska14, Flock15}.

Despite the wealth of theoretical predictions concerning condensation fronts in disks, no direct observations of fast dust growth near such fronts have been carried out. The small angular separation and emitting areas of such zones, even in the nearest protoplanetary disks, make such observations challenging. ALMA, with its unprecedented spatial resolution and sensitivity, is changing this situation dramatically. Young, actively accreting objects are good initial targets because their luminosities and disk temperature profiles move the condensation fronts out to larger, resolvable, distances \citep{Kennedy08, Blake15}. 

HL Tau, a young star in the Taurus molecular cloud ($\sim$10$^5$\,yr), was recently observed by ALMA at 
2.9, 1.3 and 0.87\,mm with a spatial resolution as good as 3\,AU (\citealt{Partnership15}, APR henceforth). Here we propose that the dark concentric rings observed in the (sub)mm interferometry of HL Tau are due to fast pebble growth near condensation fronts. We first characterize the location and width of surface brightness dips in the ALMA images before comparing their location with the expected condensation fronts in the mid-plane of the disk. We then examine the dust properties in the dips to determine if they are consistent with pebble growth.

\section{Observations}
\label{sec:obs}

The ALMA Band 3, 6 and 7 (or wavelengths of 2.9, 1.3 and 0.87\,mm, respectively) HL Tau observations were carried out in 2014, as part of its long-baseline commissioning and science verification (SV) program. With longest baselines of 15.2\,km, the HL Tau observations achieved exceptional spatial resolution, with beam sizes of 11.9$\times$8.6\,AU, 4.9$\times$3.1\,AU  and 4.2$\times$2.7\,AU at 2.9, 1.3 and 0.87\,mm (for an adopted distance of 140\,pc). More detailed descriptions of the data and calibration can be found in APR, who demonstrated the HL Tau continuum images are $\sim$axisymmetric and well fit with a series of elliptic rings. Therefore, a first order approximation of the HL Tau surface brightness is a single parameter function of the distance from the star. 

For our purposes, it is much easier to work on deprojected, circular images. We thus start with the calibrated measurements, and use CASA v4.3 \citep{McMullin07} to deproject the visibility data and generate synthesized images for all three bands. Assuming the peak continuum emission represents the location of the central star, we first shift the phase center of each band to that of HL Tau (by $\Delta_\alpha = -14\,{\rm mas}, \Delta_\delta\ = 196\,{\rm mas}$). We then deproject the visibility data using the best-fit inclination angle $i$ = 46.72$^\circ$ and PA = 138.02$^\circ$ from APR. 

From the deprojected visibility data in bands 6 and 7 we have also generated a spectral index ($\alpha$, where $I_\nu\propto\nu^{\alpha}$) map. This result is computed using the CLEAN task 
with nterms=2, based on the multi-frequency deconvolution algorithm developed by \citet{Rau11}.

\section{Charactering the emission dips}
\label{sec:dips}

\begin{figure}
\begin{center}
\vspace{-0.4cm}
\includegraphics[width=3.5in]{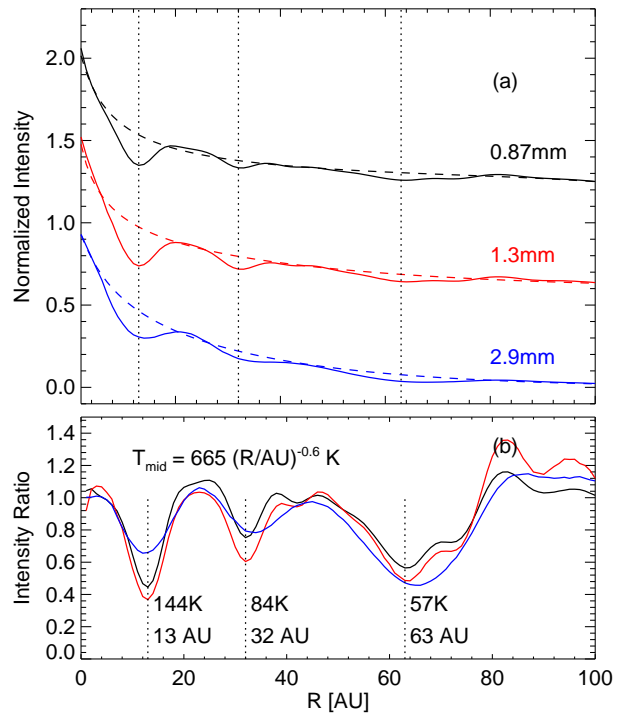}
\caption{(a) Normalized radial surface brightness distributions of HL Tau at 0.87 (blue), 1.3 (red), and 2.9\,mm (black).  The solid lines depict the observations, dashed lines the best-fit models (described in \S3). (b) The observation-to-model surface brightness ratio. The three dotted vertical lines indicate the dip minima. }
\label{fig:f_r}
\end{center}
\vspace{-0.15cm}
\end{figure}

We measure the surface brightness versus radial distance in steps of 0.5\,AU, and present the azimuthally averaged radial brightness distributions in Figure~\ref{fig:f_r}. The most consistent features in all the bands are three emission depressions, or dips, around  $\sim$13, 32 and 63\,AU. We note that there are many subtle bumps in the surface brightness distribution,   
 and we refer interested readers to a more detailed description by APR.  Here we focus on the three most prominent, $\sim$radially-symmetric dips. 

In order to constrain the dip widths and depths, a smooth fit to the surface brightness distribution is needed. We use a simple vertically isothermal model,

\begin{eqnarray}
\label{ }
I_\nu(r)& =& {\rm cos}\,i \times B_\nu (T_r) (1-e^{-\Sigma (r)  \kappa_\nu~{\rm sec}\,i})\\
T_r &=& T_0 (r/[1{\rm AU}])^q   \\
\Sigma (r) &=& \Sigma_c (r/r_c)^{-\gamma} {\rm exp}[-(r/r_c)^{2-\gamma}]
\end{eqnarray}
where $I_{\nu}$ is the surface brightness (Jy arcsec$^{-2}$), $i$ the inclination angle, $\kappa_\nu$ the continuum opacity in cm$^2$ g$^{-1}$, and $\Sigma (r)$ the dust+gas mass surface density distribution for a steady accretion disk \citep{Pringle81, Andrews09}.  The cos\,$i$ factor in eq.\,(1) accounts for the effects of the deprojection on the synthesized beam. 

Since our goal is to produce a smooth fit to the surface brightness distribution rather than a realistic physical model, we use
$\kappa_\nu$ = 0.01\cmg$(\nu/230 \rm[GHz])$ throughout the disk \citep{Ossenkopf94}.

The results of the best-fit models are plotted as dashed lines in Figure~\ref{fig:f_r}.
We then divide the observed $I_{\nu}(r)$ with the best-fit model in each band to constrain the shape of the dips, see Figure~\ref{fig:f_r} (b). The locations are consistent with those of the dark rings reported in APR.  The 13\,AU and 32\,AU intensity dips appear to be highly symmetric. Gaussian fits to the normalized radial brightness profile are listed in Table~\ref{tab:dips}. The 63\,AU dip is asymmetric, so we only list the radius of the minimum and the dip FWHM. The FWHM of the 13 and 32\,AU dips are $5-7$\,AU, about 2$\times$ the minor axis of the synthesized beams at 1.3 and 0.87\,mm, indicating that the dips are spatially resolved. The 63\,AU dip is resolved at all bands, with a FWHM of 17-18\,AU. The minimum surface intensity of the dips range from 37\% to 78\% that of the smooth continuum from the best-fit models.

\begin{deluxetable}{ccccc}
\tabletypesize{\scriptsize}
\tablecaption{Parameters of the dips, normalized to the local continuum \label{tab:dips}}
\tablewidth{0pt}
\tablehead{
\colhead{band}&\colhead{ind}&\colhead{Center [AU]} &\colhead{FWHM [AU]} &\colhead{Amplitude} 
}
\startdata

 band 7&   1&12.6$\pm$0.1    &7.2$\pm$0.3   &0.56$\pm$0.02\\
  3.0$\times$5.5\,AU &  2&32.2$\pm$0.4  &  5.3$\pm$1.0  & 0.26$\pm$0.05\\
         &3&63.5    &     17.7    &      0.44\\
        \hline
 band 6& 1&12.7$\pm$0.2   & 7.4$\pm$0.3&   0.66$\pm$0.03\\
 3.3$\times$6.6\,AU &2&32.2$\pm$0.3   &6.9$\pm$0.6 &  0.41$\pm$0.04\\
     &  3&    64.0      &   18.0      &    0.52\\
     \hline
 band 3& 1 &13.6$\pm$0.3  & 10.4$\pm$0.4  & 0.39$\pm$0.03\\
 9.1$\times$15.1\,AU &2&32.4$\pm$0.6  & 11.1$\pm$1.5 &  0.22$\pm$0.03\\
     &  3&           66.0      &   16.8      &    0.45
\enddata

\end{deluxetable}

\section{Condensation fronts of major volatiles in protoplanetary disks}

To check if the dark rings/dips in the ALMA HL Tau images are correlated with condensation fronts, we calculate the condensation temperatures of major ice species under realistic pressure ranges and compare them with the expected disk mid-plane temperature distribution. 

Here we consider the major condensible carriers of these elements based on the composition of comets, which are believed to be the best representatives of primordial icy materials in the Solar Nebula \citep{Mumma11}. We include N$_2$, one of the major carriers of N as predicted by chemical models of disks \citep{Schwarz14}.  As a homonuclear diatomic, N$_2$ cannot be directly measured in cometary comae. 

For ``pure''(single component) ices we define the condensation temperature of species $i$ as that where the thermal desorption and accretion rates are equal, i.e., 
\begin{equation}
\label{ }
n_{\rm ice}^i\times k^i_{\rm desorp} = n_{\rm gas}^i\times k^i_{\rm accr}~~~,
\end{equation}
where $k^i_{\rm desorp}$ and $k^i_{\rm accr}$ are calculated using the treatment outlined in \citet{Woitke09} and \citet{Walsh10}.

Besides pure ices, we also consider clathrate hydrates, special forms of 
crystalline water ice in which gaseous molecules can be trapped inside lattice cages of the ice.  At typical mid-plane disk pressures, abundant volatiles such as methane, H$_2$S, CO and N$_2$ are expected to be trapped in the form of clathrate hydrates before they condense as pure ice species, given a sufficiently large water ice surface area to which the vapor has access \citep{Lunine85}. This interesting feature of clathrate hydrates has been employed to explain the low N/O elemental ratio observed in comets as compared to that of the Sun's photosphere \citep{Iro03}, and the existence of methane in Titan's atmosphere \citep{Lewis71}.  

\begin{deluxetable}{lllcc}
\tabletypesize{\scriptsize}
\tablecaption{Condensation temperatures of the major volatiles in disks \label{tab:condense}}
\tablewidth{0pt}
\tablehead{
\colhead{Species}&\colhead{T$_{\rm cond}^{a}$}&\colhead{E$_{\rm b}$} &\colhead{ Cometary Abundance} &\colhead{Ref} \\
                                &\colhead{(K)}&\colhead{(K)} &\colhead{\% of H$_2$O} &
}
\startdata
H$_2$O                     &  128\,-\,155      & 5165              &100         &   1,\,5     \\
CO                              &  23\,-\,28         & 890                & 0.4\,-\,30    &  1,\,5      \\
CO$_2$	                 & 60\,-\,72          & 2605              & 2\,-\,30      &    1,\,5      \\
CH$_4$                    & 26\,-\,32          & 1000               &0.4\,-\,1.6   &   2,\,5       \\
CH$_3$OH             & 94\,-\,110        & 4355               &0.2\,-\,7      &  1,\,5        \\
N$_2$                      & 12\,-\,15          & 520                 & \nodata  &    2,\,5     \\
NH$_3$                  & 74\,-\,86           & 2965              & 0.2\,-\,1.4    &   1,\,5        \\
HCN                        & 100\,-\,120         & 4170             &0.1\,-\,0.6     &  3,\,5         \\
H$_2$S                  &45\,-\,52            & 1800            &0.1\,-\,0.6     &    4,\,5      \\
\vspace{-0.2cm}\\
\hline
\vspace{-0.18cm}\\
NH$_3\cdot$H$_2$O &78\,-\,81&\nodata& \nodata &6\\
H$_2$S$^\star$   & 77\,-\,80 &\nodata& \nodata &6\\
CH$_4^\star$ & 55\,-\,56 (69-72) &\nodata& \nodata & 6,7\\
CO$^\star$          & 45\,-\,46 (58-61) & \nodata & \nodata &6,7\\
N$_2$$^\star$           & 41\,-\,43 (55-57)& \nodata & \nodata &6
\enddata
\tablecomments{ a. Condensation temperature ranges for ices corresponding to gas number densities of 10$^{10}$- 10$^{13}$cm$^{-3}$, suitable for disk mid-planes. 
* Condensation temperatures of  clathrate hydrate formed from hexagonal ice or hydrate under gas number densities of 10$^{12}$- 10$^{13}$cm$^{-3}$. The values in parentheses are for clathrates formed from amorphous ice. \\
References: (1) \citet{Martin14}, (2) \citet{Luna14}, (3) \citet{Sandford93}, (4) \citet{Hasegawa93}, (5) \citet{Mumma11}, (6) \citet{Iro03}, (7) \citet{Lunine85} }

\end{deluxetable}

A summary of the condensation temperatures of pure ice species from such calculations is shown in Table~\ref{tab:condense}, along with the condensation temperatures of clathrate hydrates from \citet{Lunine85} and \citet{Iro03}. Our condensation temperatures are consistent with the results of \citet{Pollack91} over the same pressure range. Table~\ref{tab:condense} shows that most of the clathrate hydrates can be created at higher temperatures than their pure ice condensates. The exceptions are NH$_3$, which has a very similar temperature for its hydrate and pure ice forms, and CO$_2$, which condenses as a pure frost at higher temperatures than for which the clathrate is stable.

\begin{figure}
\begin{center}
\includegraphics[width=3.3in]{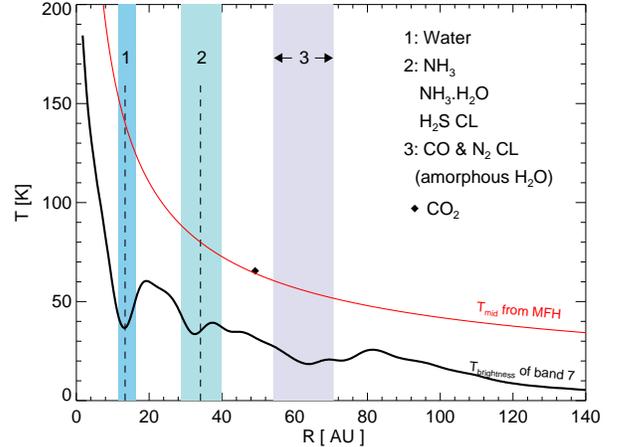}
\vspace{-0.3cm}
\caption{The expected condensation fronts in the disk mid-plane of  HL Tau. The shaded areas show the clathrate-hydrate (abbreviated as CL in the legend) condensation temperature ranges given in Table~\ref{tab:condense}, the vertical lines the mean condensation front radii.  The thick solid curve is the brightness temperature distribution of the 0.87\,mm image, the thin curve the mid-plane temperature from MFH.   A black diamond denotes the condensation temperature of pure CO$_2$ frost, which is close to edge of the 63\,AU dip. } 
\label{fig:condes}
\end{center}
\vspace{-0.1cm}
\end{figure}

In Figure~\ref{fig:condes} we compare the dip radii with the expected disk condensation front locations. We also plot the observed brightness profile, which provides a rigorous lower bound to the mid-plane dust temperature. This profile would yield a water ice line at $<$4-5 AU, unresolved by the ALMA SV data, with the NH$_3$/ammonia-hydrate condensation front near the 12 AU dip.

The actual physical temperature in this embedded disk should be larger, however, especially in the case of optically thin dust emission or significant grain coagulation.  As a more detailed estimate of the HL Tau dust temperature structure we follow \citet{Menshchikov99} (MFH henceforth), who used two-dimensional radiative transfer models to quantitatively match the available spectral energy distribution, intensity, and linear polarization data on HL Tau from near IR to (sub)mm wavelengths. Their best-fit model results in T$_{\rm mid}$ = 665 $(r/{\rm AU})^{-0.6}$\,K for the mid-plane temperature distribution.  This is greater than that estimated from the observed brightness distribution in the ALMA image, at all radii.
Interestingly, as Figure~\ref{fig:condes} shows the dips overlap nicely with the condensation fronts of the most abundant volatiles using the MFH profile: pure water condenses around the 13\,AU dip, while pure NH$_3$ or ammonia (and hydrogen sulfide) hydrates condense around the 32\,AU dip. At further distances, the condensation front of pure CO$_2$ and the onset of the CO and N$_2$ clathrate hydrate stability fields, from an amorphous water ice seed occurring near the 63\,AU dip.

\section{Dust properties inside the dips}
\label{sec:alpha}

\begin{figure*}
\begin{center}
\includegraphics[width=5.5in]{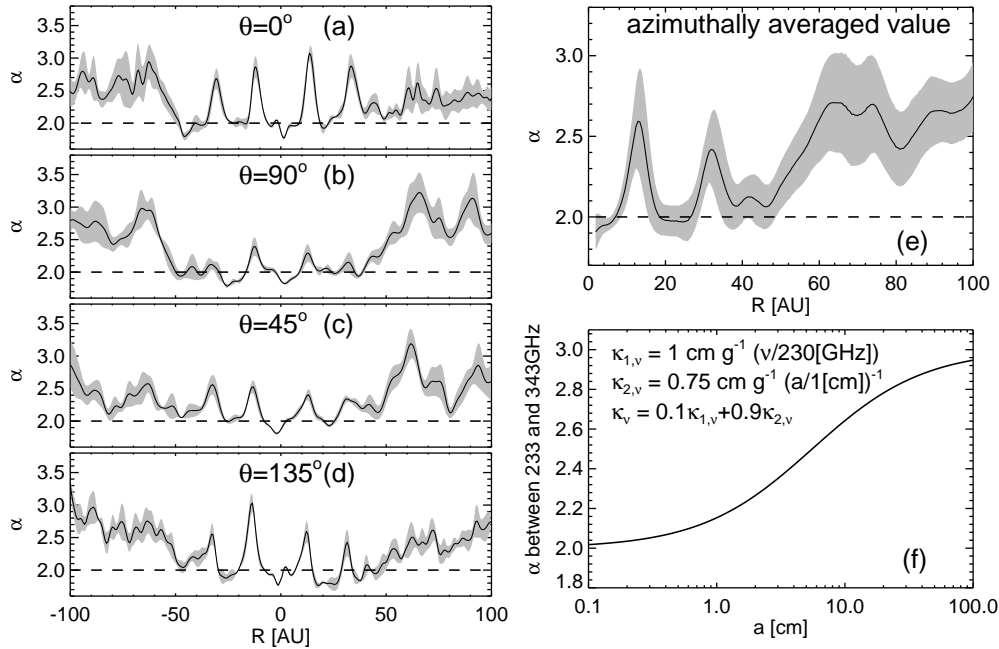}
\vspace{-0.3cm}
\caption{Panels (a)-(d): crosscuts of the spectral index $\alpha$, measured between bands 6 and 7, at four angles ($\theta$ starts from the west). The grey regions depict 1$\sigma$ uncertainty intervals. Panel (e): An azimuthal average of the continuum spectral index $\alpha$. The grey region again shows the 1\,$\sigma$ uncertainty range.  Panel (f): Spectral index $\alpha$ as a function of average dust size, $\bar{a}$, in the large dust aggregate population.}
\label{fig:alpha}
\end{center}
\end{figure*}

We have shown above that the most prominent dips in the HL Tau images are remarkably close to the expected condensation fronts of abundant volatiles. We now investigate the dust properties across these radii. 

At (sub)mm wavelengths, the spectral index $\alpha$ of the surface brightness (where $I_\nu\propto\nu^\alpha$) is a widely-used observable to characterize dust properties in protoplanetary disks \citep{Natta07}. For optically thin dust emission, $\alpha$ can be expressed as $\alpha$=2+$\beta$, where $\kappa_{\nu}\propto(\nu/\nu_0)^\beta$. The value of $\beta$ is sensitive to  the maximum dust size, $a_{\rm max}$, for a  given a size distribution, $n(a)\propto a^{q}$. If the emission is optically thick, however, $\alpha$=2 (the blackbody limit) and information on the dust size is lost.

APR provided a cross-cut  of $\alpha$ along the major axis of the as-observed image; but it is not clear if this cross-cut is representative of the full disk. In Figure~\ref{fig:alpha}, we show the radial distribution of $\alpha$, measured between 1.3 and 0.87\,mm, in crosscuts at $\theta$ = 0, 45, 90 and 135$^{\circ}$ (starting from west) along with the azimuthally averaged $\alpha (r)$. Despite large azimuthal variations, there is a clear general pattern: inside of 48\,AU $\alpha$$\sim$2, except for increases to values of $\sim$2.6 and $\sim$2.4 at the 13\,AU and 32\,AU dips, respectively. Outside 48\,AU, $\alpha$ gradually increases from 2 to 2.7 and then becomes relatively flat, except for two shallow decreases around 70 and 80\,AU near local maxima in the surface brightness.

One way to explain $\alpha$$\sim$2 invokes optically thick dust emission in the inner disk, while that in first two dips is optically thin with $\alpha$=$\beta$+2. This explanation faces two difficulties. First, a massive disk is needed to create optically thick mm-continuum with modest grain growth. Given a typical continuum opacity of  $\kappa_{\nu}$=0.01\,\cmg $(\nu/[230 \rm{GHz}])$, a surface density of $\Sigma_{\rm gas}$=100\,\gcm~ at 48\,AU is needed
to reach $\tau_{\rm 1.3\,mm}$ = 1 -- a value 20$\times$ larger than that for the MMSN \citep{Hayashi81}. 
The second issue is the large discrepancy between the temperature derived from SED fits (the MFH model) and the observed brightness temperature (Figure~\ref{fig:condes}). These should be nearly equal if the mm-dust emission is optically thick.

An alternative explanation for $\alpha$$\sim$2 that mitigates these difficulties is a scenario where the dust emission for $R\lesssim$48\,AU is effectively optically thin, but with $\beta$$\sim0$. This can occur when dust grains grow to sizes ($a$) larger than the observational wavelength ($\lambda$).  Under such conditions, the absorption cross section simply equals the particle geometric cross section -- and is thus independent of wavelength. Such significant dust growth is likely in the dense inner disk, considering that grains beyond 48\,AU have grown to $\geq$mm sizes ($\alpha$=2.77$\pm$0.13 (APR)). 
This dust growth scenario, however, also faces a problem: for a single dust size-distribution, $\alpha>2$ within the 13/32 AU dips demands smaller dust (and reduced mass surface densities) at these radii. It is difficult to understand why the disk inside of and, particularly beyond, the 13 and 32\,AU dips should have experienced more extensive dust growth than the dense inner regions of the disk.  

We suggest the apparent contradiction in the dust growth scenario can be solved by a model with two dust populations.  This model differs from the commonly invoked power law size distribution but is  commonly encountered in numerical simulations of dust growth (e.g. \citealt{Dullemond05}, \citealt{Drazkowska14}).  Our first dust population has intermediate growth,
while the second population has grown to an average size for which $a \gg \lambda$. Thus:
\begin{eqnarray}
\kappa_{\rm dust1,\nu} & = & \kappa_1 (\frac{\nu}{\nu_0})^{\beta_1} \\
\kappa_{\rm dust2,\nu} & = & \frac{\pi\bar{a}^2}{4/3\pi \rho \bar{a}^2} = \frac{3}{4\rho}\frac{1}{\bar{a}}\\
\kappa_{\rm dust,\nu} &=& (1-f)\kappa_{\rm dust1,\nu}+f\kappa_{\rm dust2,\nu}
\end{eqnarray} 
where  $\bar{a}$ and $f$ are the average dust size and mass fraction of the second dust population, and $\rho$ is the physical density of the dust.

A plausible order of magnitude estimate can be found if we assume $\kappa_{\rm dust1,\nu}$ = 1\,\cmg $(\nu/[230 \rm{GHz}])$, or dust grain growth to mm size. 
For the second population, we assume $\rho$=1\,g cm$^{-3}$, that is, ice dominated grains, and $\kappa_{\rm dust2,\nu}$ =0.75$(\bar{a}/[\rm cm])^{-1}$\cmg. We further adopt $f$=0.9, i.e., 90\% of the dust mass in the second dust population. 
For modest optical depths in the first population,the overall value of $\alpha$ between 1.3 and 0.87\,mm changes with $\bar{a}$ (Figure~\ref{fig:alpha}\,f). 
For $\bar{a}$ near 1\,cm, the cm-sized dust dominates the opacity and leads to $\alpha$$\sim2$. When $\bar{a}$ grows to decimeter-size, the mm-sized grains (population 1) dominate the mass specific dust opacity, even though they contain a small fraction of the dust+ice mass, and $\alpha$ becomes significantly greater than 2.

Thus, with significant dust aggregation in the second population, the flux and $\alpha$ behavior in the dips can be explained.  With modest mass surface densities,the fractional area of a given column of the disk covered by the dust becomes $<$unity, 
and the brightness temperature drops below the physical temperature even though individual dust aggregates are optically thick. Further (sub)mm to cm high resolution continuum images can distinguish between these two dust/disk structure scenarios through more accurate constraints on the $\alpha (r)$ distribution, as can direct measurements of the vertical and radial gas temperature distributions via molecular emission.
 
\section{Discussion}
\label{sec:dis}

The analysis presented above demonstrates that the location of the three most prominent dips in the HL Tau (sub)mm interferometric images are coincident with the expected condensation fronts of the main volatiles, using the mid-plane temperature distribution derived from observations. Further, the spectral index variation inside the dips can be explained by a bimodal dust size distribution model without the need to invoke significant surface mass density depletions.

 From our simple order of magnitude model, we infer that most of the dust mass needs to reside in a population that has grown to decimeter size scales inside the 13 and 32\,AU dips. \citet{Ros13} showed that near the water condensation front dust growth from millimeter to at least decimeter-sized pebbles is possible on a time scale of only 1000 years. This rapid growth is consistent with young age of HL Tau, between 0.1 and 1\,Myr \citep{Beckwith90, Robitaille07}.  

To date, the numerical simulations for dust growth around condensation fronts have considered only water. It is unclear if pebble growth is sufficiently rapid around other condensation fronts to explain the HL Tau results. The extra solid material delivered by clathrate hydrates cannot significantly enhance the solid mass surface density as does H$_2$O at the water snowline, since each host molecule needs $\sim$ six water molecules to form the cage structure. 
The condensation fronts of pure CO$_2$ and CO are potentially more important for enhancing mass surface density since they account for  $\sim$30\% of water abundance in comets. Nevertheless,  the model used by \citet{Ros13} can in principle to be applied to other condensation fronts.  Clathrate hydrate formation, perhaps driven by the transient warming of amorphous ice \citep{Blake91} 
via accretion bursts, can alter the ice rheology. Changes to the sticking efficiency, porosity and compaction of dust+ice aggregates may thus be central to triggering rapid pebble growth in the outer disk. 

Once a critical decimeter-sized pebble population is formed, streaming instabilities can drive the creation of $>$km-sized planetesimals \citep{Johansen14}. If fast pebble growth does preferentially occur around the condensation fronts of abundant volatiles, this would suggest that snow lines regulate the formation and chemical composition of 1-100 km planetesimals, and ultimately the formation and bulk composition of planets. 

As the disk evolves the location of condensation fronts will shift inward, perhaps countered by the effects of episodic accretion.   How this might affect dust evolution is uncertain \citep{Hubbard14}. Nonetheless, the possibility that disks such as that encircling HL Tau might be seeded with pebbles suggests that planetesimal formation might occur during early evolutionary stages. This is consistent with the cosmochemical record in our Solar system which shows that large and differentiated bodies had already begun to form $\lesssim$1\,Myr after the condensation of calcium-aluminum-rich inclusions (CAIs), the oldest minerals in the Solar system \citep{Qin08,Kleine09,Kruijer14}. 

\vspace{0.5cm}

This paper makes use of the following ALMA data sets: ADS/JAO.ALMA\#2011.0.00015.SV. ALMA is a partnership of ESO (representing its member states), NSF (USA) and NINS (Japan), together with NRC (Canada), NSC and ASIAA (Taiwan), and KASI (Republic of Korea), in cooperation with the Republic of Chile. The Joint ALMA Observatory is operated by ESO, AUI/NRAO and NAOJ.  The authors gratefully acknowledge support provided by the NSF Astronomy \& Astrophysics, NSF INSPIRE (AST-1344133), and NASA Origins of Solar Systems  grant programs.

\bibliographystyle{apj}
\bibliography{ms}

\begin{thebibliography}{}
\expandafter\ifx\csname natexlab\endcsname\relax\def\natexlab#1{#1}\fi

\bibitem[{Andrews {et~al.}(2009)Andrews, Wilner, Hughes, Qi, \&
  Dullemond}]{Andrews09}
Andrews, S.~M., Wilner, D.~J., Hughes, A.~M., Qi, C., \& Dullemond, C.~P. 2009,
  \apj, 700, 1502

\bibitem[{Beckwith {et~al.}(1990)Beckwith, Sargent, Chini, \&
  Guesten}]{Beckwith90}
Beckwith, S. V.~W., Sargent, A.~I., Chini, R.~S., \& Guesten, R. 1990, \apj,
  99, 924

\bibitem[{{Bitsch} {et~al.}(2014){Bitsch}, {Morbidelli}, {Lega}, {Kretke}, \&
  {Crida}}]{Bitsch14}
{Bitsch}, B., {Morbidelli}, A., {Lega}, E., {Kretke}, K., \& {Crida}, A. 2014,
  \aap, 570, A75

\bibitem[{{Blake} {et~al.}(1991){Blake}, {Allamandola}, {Sandford}, {Hudgins},
  \& {Freund}}]{Blake91}
{Blake}, D., {Allamandola}, L., {Sandford}, S., {Hudgins}, D., \& {Freund}, F.
  1991, Science, 254, 548

\bibitem[{{Blake} \& {Bergin}(2015)}]{Blake15}
{Blake}, G.~A., \& {Bergin}, E.~A. 2015, \nat, 520, 161

\bibitem[{{Brauer} {et~al.}(2008){Brauer}, {Henning}, \&
  {Dullemond}}]{Brauer08}
{Brauer}, F., {Henning}, T., \& {Dullemond}, C.~P. 2008, \aap, 487, L1

\bibitem[{{Desch}(2007)}]{Desch07}
{Desch}, S.~J. 2007, \apj, 671, 878

\bibitem[{{Dr{\c a}{\.z}kowska} \& {Dullemond}(2014)}]{Drazkowska14}
{Dr{\c a}{\.z}kowska}, J., \& {Dullemond}, C.~P. 2014, \aap, 572, A78

\bibitem[{{Dullemond} \& {Dominik}(2005)}]{Dullemond05}
{Dullemond}, C.~P., \& {Dominik}, C. 2005, \aap, 434, 971

\bibitem[{{Flock} {et~al.}(2015){Flock}, {Ruge}, {Dzyurkevich}, {Henning},
  {Klahr}, \& {Wolf}}]{Flock15}
{Flock}, M., {Ruge}, J.~P., {Dzyurkevich}, N., {et~al.} 2015, \aap, 574, A68

\bibitem[{{G{\"u}ttler} {et~al.}(2010){G{\"u}ttler}, {Blum}, {Zsom}, {Ormel},
  \& {Dullemond}}]{Guttler10}
{G{\"u}ttler}, C., {Blum}, J., {Zsom}, A., {Ormel}, C.~W., \& {Dullemond},
  C.~P. 2010, \aap, 513, A56

\bibitem[{{Hasegawa} \& {Herbst}(1993)}]{Hasegawa93}
{Hasegawa}, T.~I., \& {Herbst}, E. 1993, \mnras, 261, 83

\bibitem[{Hayashi(1981)}]{Hayashi81}
Hayashi, C. 1981, Progress of Theoretical Physics Supplement, 70, 35

\bibitem[{{Hubbard} \& {Ebel}(2014)}]{Hubbard14}
{Hubbard}, A., \& {Ebel}, D.~S. 2014, Icarus, 237, 84

\bibitem[{{Iro} {et~al.}(2003){Iro}, {Gautier}, {Hersant},
  {Bockel{\'e}e-Morvan}, \& {Lunine}}]{Iro03}
{Iro}, N., {Gautier}, D., {Hersant}, F., {Bockel{\'e}e-Morvan}, D., \&
  {Lunine}, J.~I. 2003, Icarus, 161, 511

\bibitem[{{Johansen} {et~al.}(2014){Johansen}, {Blum}, {Tanaka}, {Ormel},
  {Bizzarro}, \& {Rickman}}]{Johansen14}
{Johansen}, A., {Blum}, J., {Tanaka}, H., {et~al.} 2014, Protostars and Planets
  VI, 547

\bibitem[{{Kelling} {et~al.}(2014){Kelling}, {Wurm}, \&
  {K{\"o}ster}}]{Kelling14}
{Kelling}, T., {Wurm}, G., \& {K{\"o}ster}, M. 2014, \apj, 783, 111

\bibitem[{{Kennedy} \& {Kenyon}(2008)}]{Kennedy08}
{Kennedy}, G.~M., \& {Kenyon}, S.~J. 2008, \apj, 673, 502

\bibitem[{{Kleine} {et~al.}(2009){Kleine}, {Touboul}, {Bourdon}, {Nimmo},
  {Mezger}, {Palme}, {Jacobsen}, {Yin}, \& {Halliday}}]{Kleine09}
{Kleine}, T., {Touboul}, M., {Bourdon}, B., {et~al.} 2009, \gca, 73, 5150

\bibitem[{{Kretke} \& {Lin}(2007)}]{Kretke07}
{Kretke}, K.~A., \& {Lin}, D.~N.~C. 2007, \apjl, 664, L55

\bibitem[{{Kruijer} {et~al.}(2014){Kruijer}, {Touboul}, {Fischer-G{\"o}dde},
  {Bermingham}, {Walker}, \& {Kleine}}]{Kruijer14}
{Kruijer}, T.~S., {Touboul}, M., {Fischer-G{\"o}dde}, M., {et~al.} 2014,
  Science, 344, 1150

\bibitem[{{Lewis}(1971)}]{Lewis71}
{Lewis}, J.~S. 1971, Icarus, 15, 174

\bibitem[{{Luna} {et~al.}(2014){Luna}, {Satorre}, {Santonja}, \&
  {Domingo}}]{Luna14}
{Luna}, R., {Satorre}, M.~{\'A}., {Santonja}, C., \& {Domingo}, M. 2014, \aap,
  566, A27

\bibitem[{{Lunine} \& {Stevenson}(1985)}]{Lunine85}
{Lunine}, J.~I., \& {Stevenson}, D.~J. 1985, \apjs, 58, 493

\bibitem[{{Mart{\'{\i}}n-Dom{\'e}nech}
  {et~al.}(2014){Mart{\'{\i}}n-Dom{\'e}nech}, {Mu{\~n}oz Caro}, {Bueno}, \&
  {Goesmann}}]{Martin14}
{Mart{\'{\i}}n-Dom{\'e}nech}, R., {Mu{\~n}oz Caro}, G.~M., {Bueno}, J., \&
  {Goesmann}, F. 2014, \aap, 564, A8

\bibitem[{{McMullin} {et~al.}(2007){McMullin}, {Waters}, {Schiebel}, {Young},
  \& {Golap}}]{McMullin07}
{McMullin}, J.~P., {Waters}, B., {Schiebel}, D., {Young}, W., \& {Golap}, K.
  2007, in Astronomical Society of the Pacific Conference Series, Vol. 376,
  Astronomical Data Analysis Software and Systems XVI, ed. R.~A. {Shaw},
  F.~{Hill}, \& D.~J. {Bell}, 127

\bibitem[{{Men'shchikov} {et~al.}(1999){Men'shchikov}, {Henning}, \&
  {Fischer}}]{Menshchikov99}
{Men'shchikov}, A.~B., {Henning}, T., \& {Fischer}, O. 1999, \apj, 519, 257

\bibitem[{Mumma \& Charnley(2011)}]{Mumma11}
Mumma, M.~J., \& Charnley, S.~B. 2011, \araa, 49, 471

\bibitem[{{Natta} {et~al.}(2007){Natta}, {Testi}, {Calvet}, {Henning},
  {Waters}, \& {Wilner}}]{Natta07}
{Natta}, A., {Testi}, L., {Calvet}, N., {et~al.} 2007, Protostars and Planets
  V, 767

\bibitem[{{{\"O}berg} {et~al.}(2011){{\"O}berg}, {Murray-Clay}, \&
  {Bergin}}]{Oberg11}
{{\"O}berg}, K.~I., {Murray-Clay}, R., \& {Bergin}, E.~A. 2011, \apjl, 743, L16

\bibitem[{{Ossenkopf} \& {Henning}(1994)}]{Ossenkopf94}
{Ossenkopf}, V., \& {Henning}, T. 1994, \aap, 291, 943

\bibitem[{{Partnership} {et~al.}(2015){Partnership}, {Brogan}, {Perez}, \&
  {Hunter}}]{Partnership15}
{Partnership}, A., {Brogan}, C.~L., {Perez}, L.~M., \& {Hunter}, e.~a. 2015,
  ArXiv e-prints, arXiv:1503.02649

\bibitem[{{Pollack} {et~al.}(1996){Pollack}, {Hubickyj}, {Bodenheimer},
  {Lissauer}, {Podolak}, \& {Greenzweig}}]{Pollack96}
{Pollack}, J.~B., {Hubickyj}, O., {Bodenheimer}, P., {et~al.} 1996, Icarus,
  124, 62

\bibitem[{{Pollack} {et~al.}(1991){Pollack}, {Lunine}, \&
  {Tittemore}}]{Pollack91}
{Pollack}, J.~B., {Lunine}, J.~I., \& {Tittemore}, W.~C. 1991, {Origin of the
  Uranian satellites}, ed. J.~T. {Bergstralh}, E.~D. {Miner}, \& M.~S.
  {Matthews}, 469--512

\bibitem[{Pringle(1981)}]{Pringle81}
Pringle, J.~E. 1981, \araa, 19, 137

\bibitem[{{Qin} {et~al.}(2008){Qin}, {Dauphas}, {Wadhwa}, {Masarik}, \&
  {Janney}}]{Qin08}
{Qin}, L., {Dauphas}, N., {Wadhwa}, M., {Masarik}, J., \& {Janney}, P.~E. 2008,
  Earth and Planetary Science Letters, 273, 94

\bibitem[{{Rau} \& {Cornwell}(2011)}]{Rau11}
{Rau}, U., \& {Cornwell}, T.~J. 2011, \aap, 532, A71

\bibitem[{Robitaille {et~al.}(2007)Robitaille, Whitney, Indebetouw, \&
  Wood}]{Robitaille07}
Robitaille, T.~P., Whitney, B.~A., Indebetouw, R., \& Wood, K. 2007, \apjs,
  169, 328

\bibitem[{{Ros} \& {Johansen}(2013)}]{Ros13}
{Ros}, K., \& {Johansen}, A. 2013, \aap, 552, A137

\bibitem[{{Sandford} \& {Allamandola}(1993)}]{Sandford93}
{Sandford}, S.~A., \& {Allamandola}, L.~J. 1993, Icarus, 106, 478

\bibitem[{{Schwarz} \& {Bergin}(2014)}]{Schwarz14}
{Schwarz}, K.~R., \& {Bergin}, E.~A. 2014, \apj, 797, 113

\bibitem[{{Seizinger} \& {Kley}(2013)}]{Seizinger13}
{Seizinger}, A., \& {Kley}, W. 2013, \aap, 551, A65

\bibitem[{{Stevenson} \& {Lunine}(1988)}]{Stevenson88}
{Stevenson}, D.~J., \& {Lunine}, J.~I. 1988, Icarus, 75, 146

\bibitem[{{Walsh} {et~al.}(2010){Walsh}, {Millar}, \& {Nomura}}]{Walsh10}
{Walsh}, C., {Millar}, T.~J., \& {Nomura}, H. 2010, \apj, 722, 1607

\bibitem[{{Woitke} {et~al.}(2009){Woitke}, {Kamp}, \& {Thi}}]{Woitke09}
{Woitke}, P., {Kamp}, I., \& {Thi}, W.-F. 2009, \aap, 501, 383

\end{thebibliography}

\end{document}